\documentclass[pra,reprint]{revtex4-1}
\usepackage{graphicx}
\usepackage{dcolumn}
\usepackage{bm}
\usepackage{amsmath}
\usepackage{amsfonts}
\usepackage{psfrag}
\usepackage[pdftex,bookmarks,colorlinks=true,linkcolor = magenta,urlcolor  = red,citecolor = blue]{hyperref}
 \usepackage[usenames,dvipsnames]{color}

\def\bra#1{\mathinner{\langle{#1}|}} 
\def\ket#1{\mathinner{|{#1}\rangle}}

\newcommand{\Eq}[1]{Eq. (\ref{#1})}

\newcommand{\ca}[2]{c_{#1,\mathbf{#2}}}
\newcommand{\cc}[2]{c^{\dagger}_{#1, \mathbf{#2}}}

\newcommand{\aaa}[2]{a_{\mathbf{#1},#2}}
\newcommand{\aac}[2]{a^{\dagger}_{\mathbf{#1},#2}}

\newcommand{\sub}[1]{_{\boldsymbol{#1}}}

\newcommand{\subthz}[0]{_{\text{THz}}}

\begin{document}

\title{Terahertz emission from AC Stark-split asymmetric intersubband transitions}
\author{Nathan Shammah$^{1}$, Chris C. Phillips$^{2}$, and Simone \surname{De Liberato}$^{1}$}
\affiliation{$^{1}$School of Physics and Astronomy, University of Southampton, Southampton, SO17 1BJ, UK}
\affiliation{$^{2}$Physics Department, Imperial College London, London SW7 2AZ, UK}
\begin{abstract}
Transitions between the two states of an AC Stark-split doublet are forbidden in centro-symmetric systems, and thus almost impossible to observe in experiments performed with  atomic clouds. However, electrons trapped in nanoscopic heterostructures can behave as artificial atoms, with the advantage that the wavefunction symmetry can be broken by using asymmetric confining potentials. Here we develop the many-body theory describing the intra-doublet emission of a resonantly pumped intersubband transition in a doped asymmetric quantum well, showing that in such a system the intra-doublet emission can be orders of magnitude higher than in previously studied systems.
This emission channel, which lies in the terahertz range, and whose frequency depends upon the pump power, opens the way to the realization of a new class of monolithic and tunable terahertz emitters.
 \end{abstract}
\maketitle  
\section{Introduction}
\begin{figure}[t!]
\begin{center}
\includegraphics[width=8.6cm]{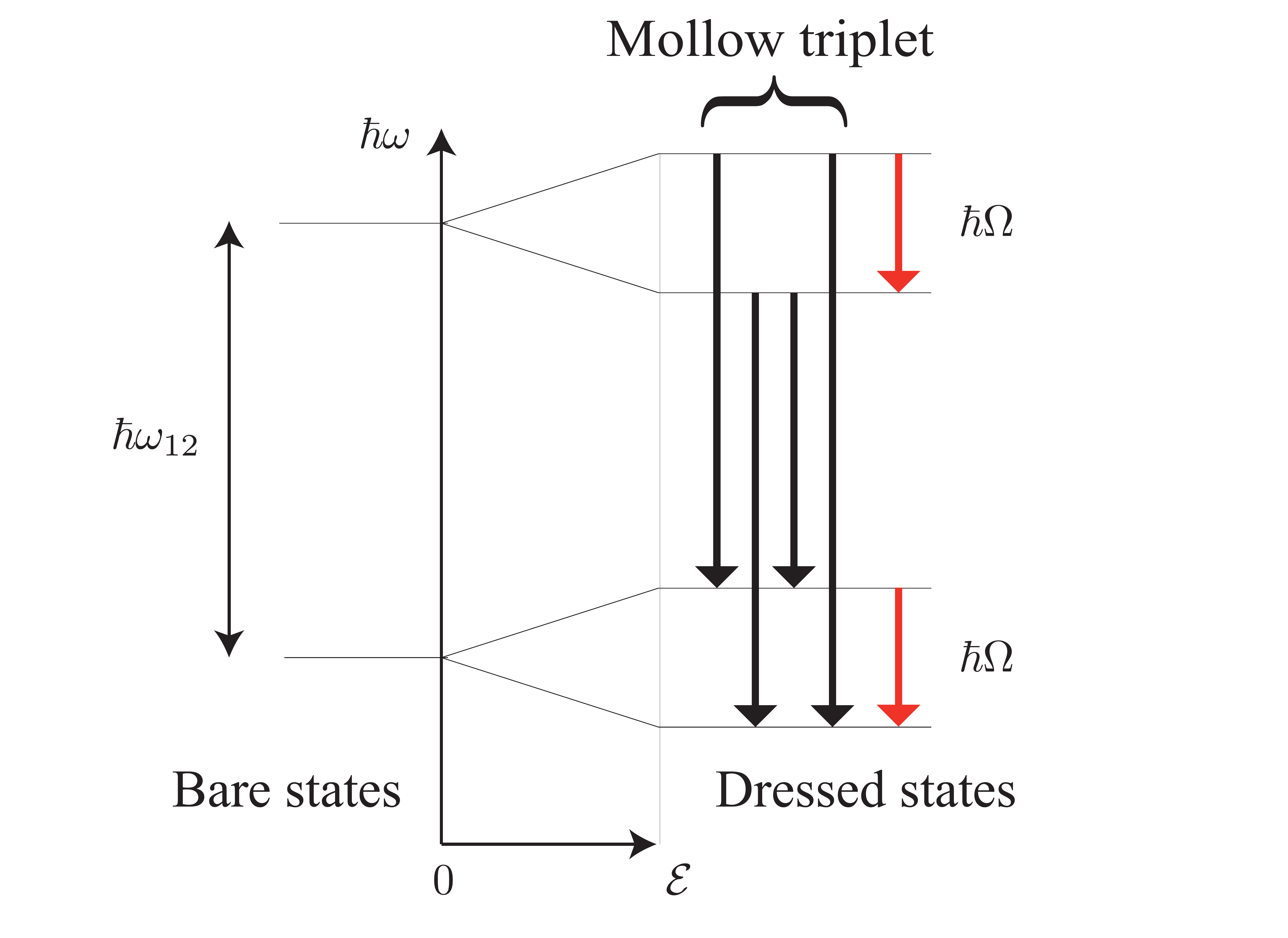}
\caption{\label{split}
Spectrum of a two level system resonantly driven with amplitude $\mathcal{E}$.
Inter-doublet transitions (thick black arrows) gives rise to the Mollow triplet with emission centred at frequencies $\omega_{12}$ and $\omega_{12}\pm\Omega)$. Intra-doublet transitions (thick red arrows) at frequency $\Omega$, are forbidden in centro-symmetric systems.}
\end{center}
\end{figure}

When an electronic transition of a quantum system is driven by a strong optical pump, the field dresses the system, splitting the energy levels into doublets (see Fig. \ref{split}), whose energy separation is given by the Rabi splitting 
\begin{eqnarray}
\label{Omega}
\hbar\Omega=d{\mathcal{E}},
\end{eqnarray}
where ${d}$ is the transition's dipole and ${\mathcal{E}}$ is the amplitude of the applied electric field. This effect is known as the dynamical (or AC) Stark effect \cite{Autler55}. The fluorescence spectrum of such a system is characterized by the Mollow triplet \cite{Mollow69,Delvalle10}, which arises from transitions between states of neighbouring Rabi doublets (thick black arrows in Fig. \ref{split}). Transitions between states belonging to the same Rabi doublet (thick red arrows in Fig. \ref{split}) are dipole-forbidden in centro-symmetric systems. 
By breaking the symmetry of the potential that confines the electrons, this selection rule can be lifted \cite{Kibis09}, allowing new emission lines centred at the Rabi frequency $\Omega$ to be observed. 

\begin{figure}[h!]
\begin{center}
\includegraphics[width=8.6cm]{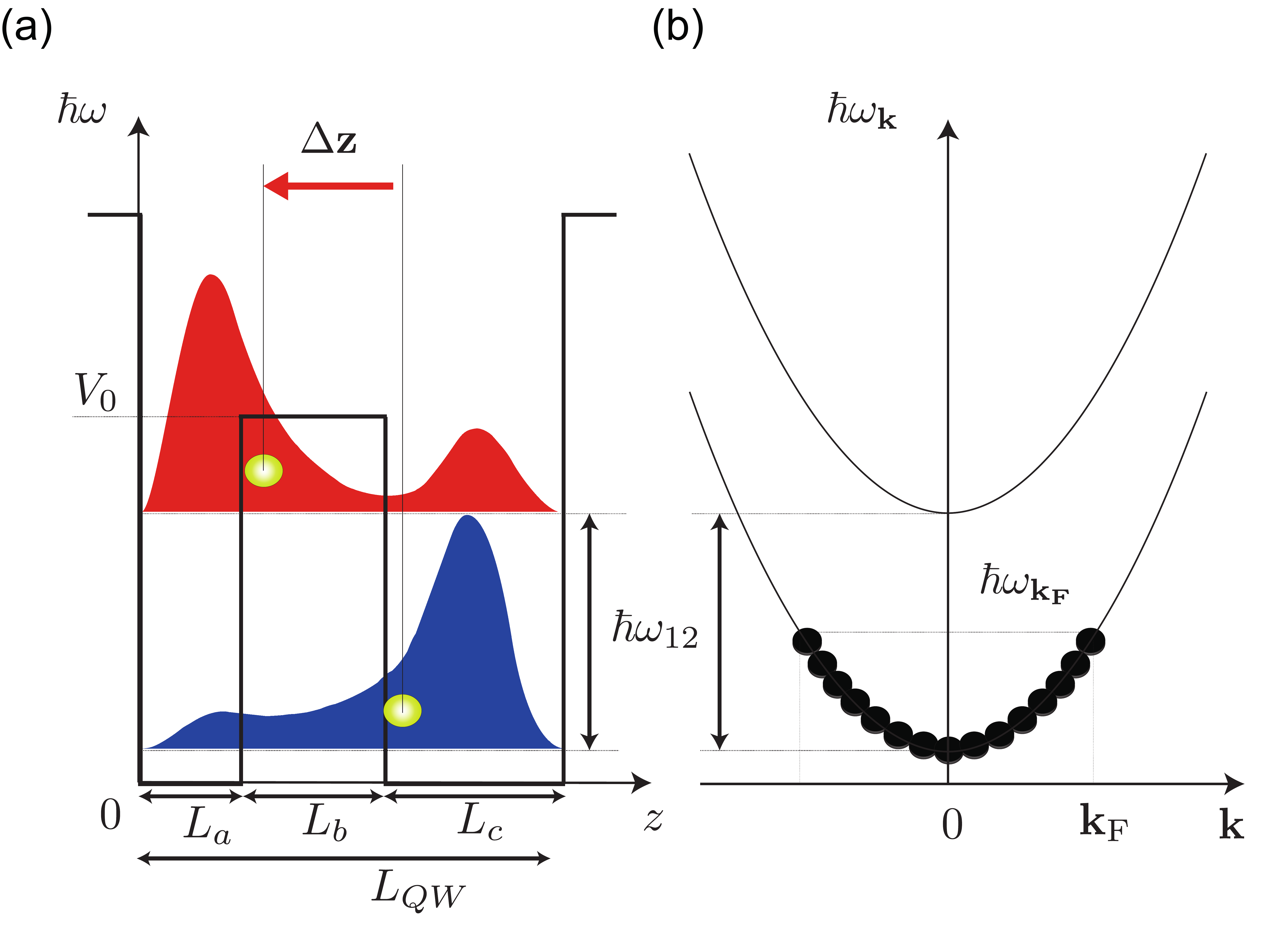}
\caption{
\label{isbt+qw}
Panel (a): An asymmetric quantum well and the wavefunctions of its first two conduction subbands. Due to the asymmetry, the average charge position in the two subbands is different. Electrons cycling between them under the effect of a resonant pump generate a radiating dipole of length $\Delta \mathbf{z}$, oscillating at the Rabi frequency $\Omega$.  
Panel (b): The parabolic dispersion in $\mathbf{k}$-space of the first two conduction subbands. }
\end{center}
\end{figure}

A more intuitive understanding of such an emission channel can be gained reasoning in the time domain, where the electrons undergo Rabi oscillations at frequency $\Omega$ between the initial and final bare states under the effect of the pump. In a centro-symmetric system, the average electron position is the same for both these states, and Rabi oscillations do not result in a net charge oscillation. If the symmetry of the electronic wavefunctions is broken instead, the electronic charge oscillates back and forth, and we expect the system to radiate as a dipole oscillating at frequency $\Omega$ (see Fig. \ref{isbt+qw}(a) for a pictorial representation).

Interband transitions in quantum dots have been proposed as candidates for observing this emission channel  \cite{Savenko12}, but the magnitude of their asymmetric dipole is small and difficult to control, as it relies on the intrinsic anisotropy of the crystal lattice. Intersubband transitions (ISBTs) in doped quantum wells (QWs) appear to be a better candidate, thanks to the possibility of tailoring their asymmetry by engineering the confining structure \cite{Geiser12}. 
Moreover, for these systems, the Rabi splitting lies in the THz domain \cite{Dynes05}, and it depends on the intensity of the pump beam, as shown by \Eq{Omega}. Symmetry-forbidden transitions in ISBTs, apart from their fundamental interest in quantum optics, could thus empower a new generation of extremely tunable, monolithic THz emitters. 

The possibility of using asymmetric QWs to obtain THz radiation was recently explored by two of the present authors \cite{DeLiberato13} in the polaritonic case, where the splitting is not due to the high intensity of the pump laser, but to the strong coupling of ISBTs to the vacuum field of a photonic microcavity. In that case, in the dilute excitation regime, the spectrum of the system is composed of quasi-bosonic excitations called intersubband polaritons \cite{Dini03,Ciuti05,Pereira07,Anappara09,DeLiberato09b,Todorov10}. In Ref. \cite{DeLiberato13}, it was shown that an asymmetric QW structure can give rise to scattering between different polaritonic branches, leading to the possibility of designing an efficient THz laser, whose emission frequency could be partially tuned by modifying the electron density in the QWs \cite{Anappara05,Gunter09}. Related works, that also use symmetry breaking to observe otherwise forbidden emissions,  have also recently been proposed \cite{Kavokin10,Savenko11,Delvalle11,Kavokin12, Ridolfo12,Tighineanu14}.

Compared with these existing works, the scheme here presented offers the advantages of an extreme tunability of the emission frequency, and of a comparatively simple and flexible design, because it does not rely on  a resonant photonic cavity coupled to the ISBT.
From a theoretical point of view, a major difference between this scheme and all the others cited above, 
is that here we are interested in the full electron dynamics, and so we cannot limit ourselves to the bosonic or quasi-bosonic regime. We will thus have to work in a fermionic basis, without any bosonization approximation. 
Moreover, in this non-bosonic, nonlinear regime, the electrons in ISBTs do not generally behave as independent dipoles (as it has been recently proved in Ref. \cite{DeLiberato13a}) and thus we cannot a priori rely on the single dipole theory developed in Ref. \cite{Kibis09}. 

In the following, we will thus develop a general theory of the spontaneous emission from the resonantly driven ISBT of a two dimensional electron gas (2DEG). After developing the general formalism to calculate the emission efficiency, we will perform a numerical study of the magnitude of the asymmetric dipole achievable in a QW structure.
This will allow us to quantify the experimentally achievable emission efficiency, and to address future experiments toward the best sample geometry.

The paper is structured as follows: in Sec. \ref{Atom} we discuss why symmetry-forbidden transitions have never been observed in atomic physics experiments. In Sec. \ref{Theory} we develop a theory describing optically pumped ISBTs. Such a theory will then be used in Sec. \ref{Num} to estimate the THz emission efficiency for a realistic asymmetric device. Finally, conclusions and perspectives are drawn in Sec. \ref{Conclusions}.

\begin{figure}[t!]
\begin{center}
\includegraphics[width=8.6cm]{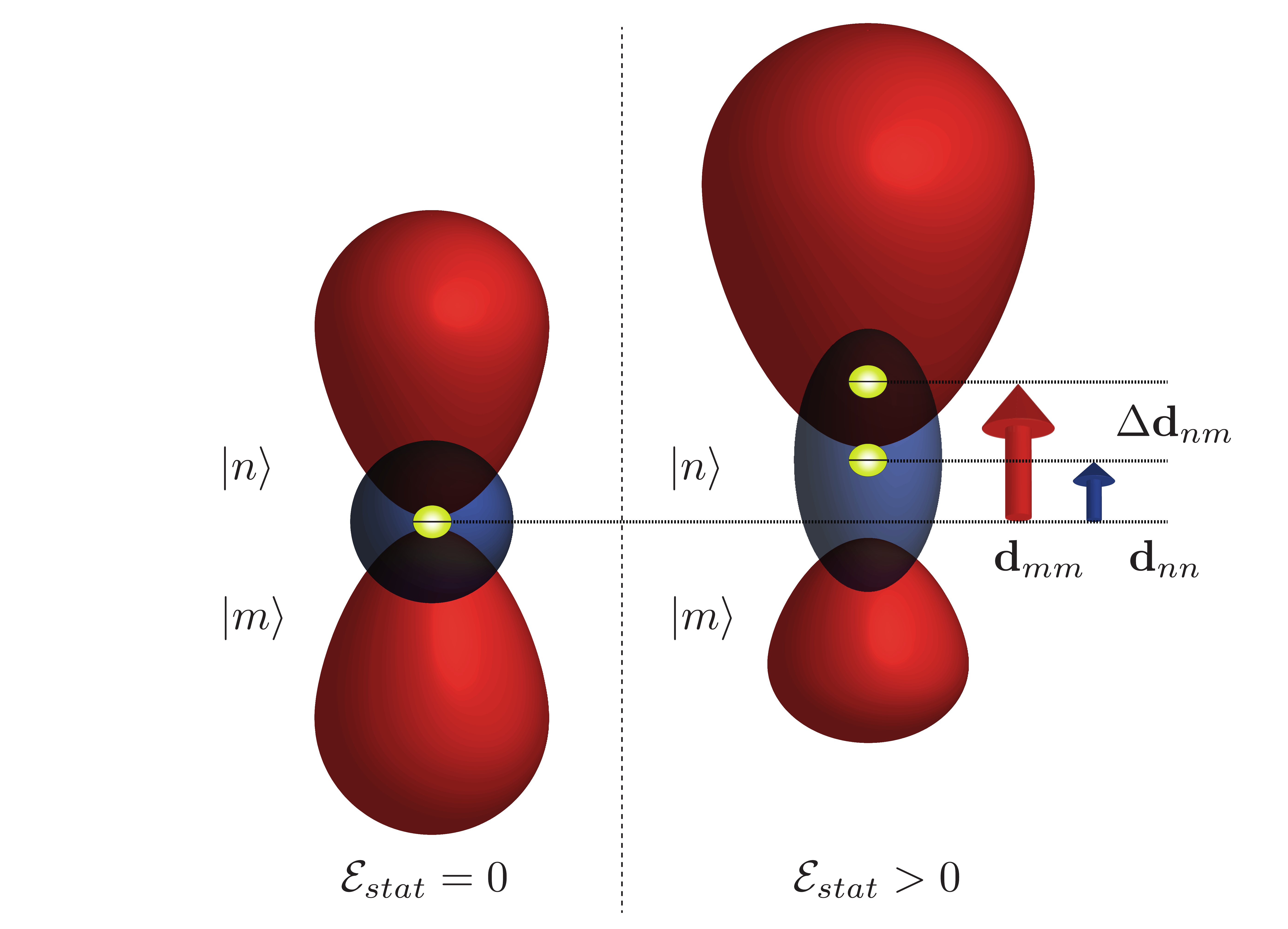}
\caption{\label{atom}
In an atom, the symmetry of the electronic orbitals can, in principle, be broken by applying an external static electric field, $\boldsymbol{\mathcal{E}_{\text{stat}}}$. As the orbitals are deformed in the same direction, the two induced dipoles, $\mathbf{d}_{nn}$ and $\mathbf{d}_{mm}$, are parallel, and the asymmetric dipole moment, $\Delta \mathbf{d}_{nm}=\bra{n}\mathbf{{d}}\ket{n}-\bra{m}\mathbf{{d}}\ket{m}$, only grows differentially.}
\end{center}
\end{figure}

\section{Asymmetric dipole in atomic systems}
\label{Atom}
Before proceeding with our study of THz emission in asymmetric QWs, it is instructive to recall why symmetry-forbidden emission channels have never been observed with atomic systems \cite{Haroche}, which are usually an ideal testbed to observe quantum optics phenomena. Although atoms are intrinsically symmetric, their symmetry can be broken applying
a static electric field \cite{Wood97}, as pictorially shown in Fig. \ref{atom}.

The coupling with an external, static electric field $\boldsymbol{\mathcal{E}_{\text{stat}}}$, can be described by the interaction Hamiltonian 
\begin{eqnarray}
V_{\text{stat}} =-\ \mathbf{{d}}\cdot\boldsymbol{\mathcal{E}_{\text{stat}}},
\label{Vdip}
\end{eqnarray}
where $\mathbf{d}$ is the atomic dipole.
To first order in perturbation, the asymmetric dipole between two states, $\ket{n}$ and $\ket{m}$,
 whose energy difference is $\hbar\omega_{nm}$, is
\begin{eqnarray}
\label{dasy}
\Delta\mathbf{d}_{nm}&=&\bra{n}\mathbf{{d}}\ket{n}-\bra{m}\mathbf{{d}}\ket{m}.
\end{eqnarray} 
The ratio between the unperturbed transition dipole 
\begin{eqnarray}
\mathbf{d}_{nm}&=&\bra{n}\mathbf{{d}}\ket{m},
\end{eqnarray}
and the asymmetric one $\Delta\mathbf{d}_{nm}$, is thus of order
$\frac{\mathbf{{d}}\cdot\boldsymbol{\mathcal{E}_{\text{stat}}}
}{\hbar\omega_{nm}}$, and, for non-ionizing
 fields, it is always much less than unity.
A better estimate can be obtained by introducing the electric polarizability for a state $\ket{n}$, $\alpha_n$, allowing the asymmetric dipole to be rewritten as
\begin{eqnarray}
\label{alfad}
\Delta\mathbf{d}_{nm}&=&\left(\alpha_n-\alpha_m\right)\boldsymbol{\mathcal{E}}_{\text{stat}}.
\end{eqnarray}
Alkali atoms possess the strongest polarizabilities of both the ground state ($n=S_{1/2}$) and of the first excited states ($m=P_{J}$, with $J=\{1/2, 3/2\}$), 
due to the fact that just one valence electron lies in the outer shell. Interpolating data from both experimental measurements and theoretical calculations \cite{Mitroy10}, one finds that the difference in the polarizabilities of the ground and first excited states of Alkali atoms: Li, Na, K, Rb, and Cs, are at most $\simeq 1000$  $a_{0}^{3}$, where $a_{0}$ is the Bohr radius. 
Given that the transition dipoles are instead 
$\approx 4$ $e a_{0}$, even considering an extremely strong electric field up to $\mathcal{E}_{stat}=10^{7}$ V/m \cite{Windholz89} in \Eq{alfad}, we obtain 
\begin{eqnarray}
\Delta d_{nm}/d_{nm}\lesssim 0.003,
\end{eqnarray}
implying that any effect due to the asymmetric dipole will be very challenging to observe.

\section{Main theory}
\label{Theory}
\subsection{ISBT under intense resonant pumping: AC Stark splitting}
In a QW, spatial confinement along the growth $z$ direction splits the electron bands into multiple, quasi-parallel subbands, as shown in Fig. \ref{isbt+qw}(b). Doping can raise the Fermi energy, $\hbar\omega_{\text{F}}$, between the first and the second subband edges, so that the first conduction subband is partially filled with a 2DEG. 
Defining $c_{j,\mathbf{k}}$ to be the annihilation operator for an electron in the $j$th subband, with in-plane wavevector $\mathbf{k}$ and energy $\hbar\omega_{j,{k}}$, we can write the Hamiltonian describing a doped quantum well under the effect of a pump which is quasi-resonant with the transition between the first two conduction subbands as
\begin{eqnarray}
H_{0}&=&\hbar\sum_{j=\{1,2\},\mathbf{k}} \omega_{j,{k}}\cc{j}{k}\ca{j}{k}\nonumber\\
&+&\frac{\hbar\Omega}{2}\sum_{\mathbf{k}}\left( \cc{2}{k+\bar{q}}\ca{1}{k}e^{i\omega_L t}+ \cc{1}{k}\ca{2}{k+\bar{q}}e^{-i\omega_L t}\right),
\label{H_0}
\end{eqnarray}
where $\omega_L$ is the pump frequency, $\bar{\mathbf{q}}$ its in-plane wavevector, and $\Omega$ the Rabi frequency as defined in \Eq{Omega}, proportional to the square root of the pump intensity.
Since the dipole of the ISBT lies parallel to the growth axis, we can write explicitly
\begin{eqnarray}
\Omega=ez_{12}\mathcal{E}/\hbar,
\label{ez12}
\end{eqnarray} 
where $e$ is the electron charge, $\mathcal{E}$ the amplitude of the pump field,
\begin{eqnarray}
z_{ij}=\int\psi^{*}_{i}(z)\psi_{j}(z)zdz,
\label{zij}
\end{eqnarray} and $\psi_{j}(z)$ is the envelope function of an electron in the $j$th subband.  

Hereafter, we neglect the photon momentum when it is compared to the electronic one, i.e., we approximate $\omega_{2,k+q}\simeq\omega_{2,k}$, where $\omega_{2,k}=\omega_{1,k}+\omega_{12}$. By choosing a suitable rotating frame, and by setting the pump resonant with the ISBT bare frequency, $\omega_L=\omega_{12}$, \Eq{H_0} can be rewritten as
\begin{eqnarray}
H'_{0}=\sum_{\mathbf{k}} H_{\mathbf{k}}=\frac{\hbar\Omega}{2}\sum_{\mathbf{k}}\left( \cc{2}{\mathbf{k+\bar{q}}}\ca{1}{\mathbf{k}}+ \cc{1}{\mathbf{k}}\ca{2}{\mathbf{k+\bar{q}}}\right).\quad
\label{H_0p}
\end{eqnarray}

Notice that, since all the interactions are spin-conserving, we have omitted the spin index in the electron operator here; all the sums over electronic wavevectors are thus implicitly assumed to also be summed over electron spin. We also neglect the electron-electron Coulomb interaction, since it has been showed to amount, to leading order, only to a renormalization of the intersubband energy $\hbar\omega_{12}$, an effect usually referred to as depolarization shift \cite{Nikonov97,Kyriienko12,DeLiberato12,Luc13}.  
In the following, we assume that the system is in a cryogenic environment, allowing us to disregard all temperature effects.

\begin{figure}[t!]
\begin{center}
\includegraphics[width=8.6cm]{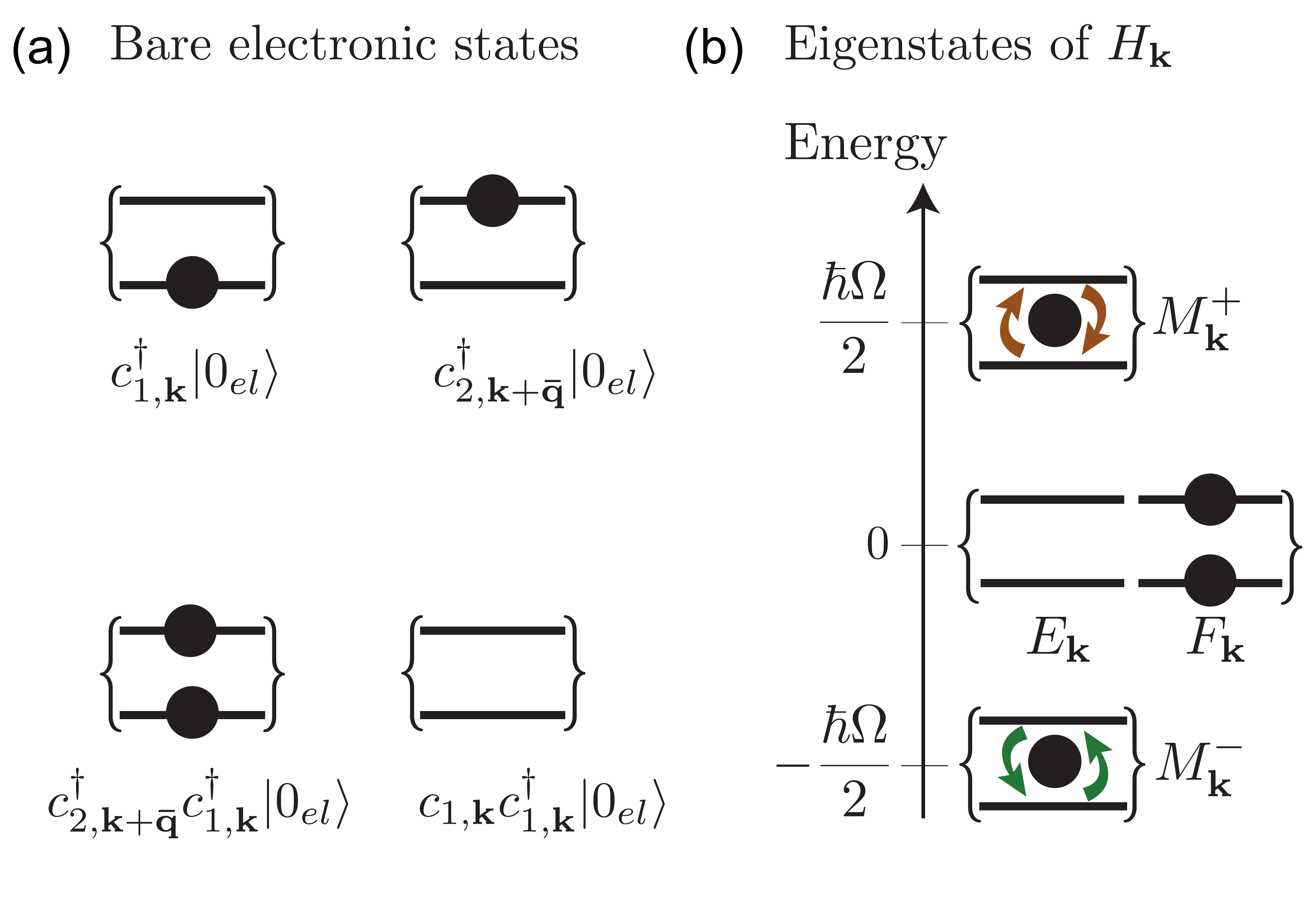}
\caption{
\label{eigenfig} Representation of the bare states (a) and eigenstates (b) of $H_{\mathbf{k}}$.}
\end{center}
\end{figure}

The operators of a pair of electrons, one in the first subband with wavevector $\mathbf{k}$, and one in the second subband 
with wavevector $\mathbf{k+\bar{q}}$, appear only once in the sum in \Eq{H_0p}, in the term $H_{\mathbf{k}}$, as shown pictorially in Fig. \ref{eigenfig}(a). Each $H_{\mathbf{k}}$ thus acts on a different 4-dimensional subspace of the electronic Hilbert space (two electronic states, each of them full or empty).  
Defining the ground state of the electronic system without the pump coupling ($\Omega=0$) as
\begin{eqnarray}
\ket{G}=\prod\limits_{|\mathbf{k}|<k_{\text{F}}}c^{\dagger}_{1,\mathbf{k}}\ket{0_{el}}, 
\end{eqnarray}
$\ket{0_{el}}$ being the empty conduction band, and $k_{\text{F}}$ the Fermi wavevector, we can
diagonalize $H_{\mathbf{k}}$, writing  its eigenvectors explicitly as
\begin{eqnarray}
\label{EF}
E\sub{k}\ket{G}&=&\ca{1}{k}\ket{G},\nonumber \\
F\sub{k}\ket{G}&=&\cc{2}{k+\bar{q}}\ket{G},\\
M^{\pm}_{\mathbf{k}}\ket{G}&=&\frac{1}{\sqrt{2}}(\cc{2}{k+\bar{q}}\ca{1}{k}\pm1)\ket{G}.\nonumber
\end{eqnarray}
A representation of these four states is presented in Fig. \ref{eigenfig}(b).
The first two states of \Eq{EF} describe the full and empty states, in which states in neither or both subbands in the considered subspace of $H_{\mathbf{k}}$ are occupied. As these states do not couple with the pump laser, in the rotating frame in which we are working, both these states are degenerate and have zero energy. 
The other states in \Eq{EF}, with energy $\pm\frac{\hbar\Omega}{2}$, are instead states where only one electron is present and, under the action of the pump, it cycles between the two subbands.

In order to be able to calculate the THz emission due to the presence of an asymmetric dipole, we will start by finding the full many-body eigenstates of \Eq{H_0p}, in order to be then able to perturbatively calculate emission using Fermi's Golden Rule. 

From the decomposition of $H'_0$ in a sum of commuting Hamiltonians in \Eq{H_0p}, a general eigenvector of 
$H'_0$ can be put in the form 
\begin{eqnarray}
\ket{\psi}&=&\prod_{\mathbf{k}\in S_+}M^+\sub{k} \prod_{\mathbf{k}\in S_-}M^-\sub{k} \prod_{\mathbf{k}\in S_F}F\sub{k}
\prod_{\mathbf{k}\in S_E}E\sub{k}\ket{G},
\label{psi} 
\end{eqnarray}
where the four sets $S_+$, $S_-$, $S_F$, and $S_E$ are a partition of the Fermi sphere, with cardinalities $N_+$, $N_-$, $N_F$, and $N_E$, respectively, which are constrained by the fact that the total electron number is conserved  
\begin{eqnarray}
N&=&N_++N_-+2N_F.
\label{Ntot}
\end{eqnarray}
The eigenvalue of a given many-body state $\ket{\psi}$ in \Eq{psi} is thus given by
\begin{eqnarray}
\label{Energy}
\hbar\omega_{0}&=&\frac{\hbar\Omega}{2}(N_+-N_-).
\end{eqnarray}

For further reference, it is useful to calculate the following operator products,
\begin{eqnarray}
\label{prod}
F_{\mathbf{k}}E_{\mathbf{k}}&=&\frac{M_{\mathbf{k}}^++M_{\mathbf{k}}^-}{\sqrt{2}},\\
F_{\mathbf{k}}M_{\mathbf{k}}^{\pm}&=&{\pm}\frac{F_{\mathbf{k}}}{\sqrt{2}},\nonumber\\
E_{\mathbf{k}}M_{\mathbf{k}}^{\pm}&=&{\pm}\frac{E_{\mathbf{k}}}{\sqrt{2}},\nonumber\\
E_{\mathbf{k}}E_{\mathbf{k}}&=&F_{\mathbf{k}}F_{\mathbf{k}}=0.\nonumber
\end{eqnarray}

\subsection{The light-matter coupling Hamiltonian}
In order to calculate the photonic emission rate from dipolar transitions between the states of the pumped electronic system, we need to couple it to the electromagnetic field continuum. This can be accomplished by considering the full Hamiltonian
\begin{eqnarray}
H=H'_0+H_{EM}+V,
\end{eqnarray}
where $H_{EM}$ is the Hamiltonian of the free electromagnetic field,
\begin{eqnarray}
H_{EM}&=&\hbar \sum_{\mathbf{q},q_z} \omega_{{q},q_z} a^{\dagger}_{\mathbf{q},q_z}a_{\mathbf{q},q_z},
\end{eqnarray}
such that $a_{\mathbf{q},q_z}$ is the bosonic operator annihilating a photon with in-plane and normal wavevectors $\mathbf{q}$ and $q_z$ respectively, and energy $\hbar \omega_{{q},q_z} $.
The term $V$ describes the coupling between the ISBTs and the electromagnetic field. In the rotating frame this reads
\begin{eqnarray}
\label{Vint}
V&=&\sum_{\mathbf{k},\mathbf{q},q_z} (\aac{-q}{q_z}+\aaa{q}{q_z}) [\sum_{j=\{1,2\}}\chi_{jj,{q},q_{z}}\cc{j}{k+q}\ca{j}{k} \label{thz}
\nonumber\\
&+&  \chi_{12,{q},q_{z}} ( e^{i\omega_{12}t}\cc{2}{k+q}\ca{1}{k}+e^{-i\omega_{12}t} \cc{1}{k}\ca{2}{k-q})],\quad
\end{eqnarray}
where the explicit expressions for the coupling coefficients can be written as
\begin{eqnarray}
\label{chi}
\chi_{ij,q,q_z}&=&\mathcal{E}_0\frac{q}{\sqrt{q^2+q_z^2}}z_{ij},
\end{eqnarray}
with 
\begin{eqnarray}
\label{zpf}
\mathcal{E}_{0}=\sqrt{\hbar\omega_{{q},q_{z}}/(2\epsilon_{0}\epsilon_{r}\mathcal{V})},
\end{eqnarray}
being the zero point fluctuation of the electromagnetic field, $\mathcal{V}$ the quantization volume, and $\epsilon_r$ the relative dielectric constant inside the QW. Notice that, because of the selection rules of ISBTs, the photonic field is assumed to be TM polarized.

An inspection of  \Eq{Vint} reveals that, for $q\neq\bar{q}$, V couples subspaces corresponding to different values of $\mathbf{k}$ in \Eq{H_0p}. Therefore, we cannot limit ourselves to treat each subspace independently and we must calculate the transition matrix elements whilst taking into account the full many-body nature of the problem. 

There are two qualitatively different kinds of terms that can be identified in \Eq{Vint}. The last two terms, whose coupling is proportional to the intersubband dipole, 
$ez_{12}$, will give an emission centred on the unperturbed frequency $\omega_{12}$, with two satellite peaks at $\omega_{12}\pm \Omega$. In the following, we disregard these Mollow-like emission components, and concentrate instead on the remaining part of the interaction Hamiltonian  
\begin{eqnarray}
V\subthz= \sum_{\mathbf{k},\mathbf{q},q_z}&&(\chi_{11,{q},q_{z}}c^{\dagger}_{1,\mathbf{k+q}}\ca{1}{k}+\chi_{22,{q},q_{z}}c^{\dagger}_{2,\mathbf{k+q}}\ca{2}{k})\nonumber\\
&& \times(\aac{-q}{q_z}+\aaa{q}{q_z}),
\label{Vthz0}
\end{eqnarray}
which is responsible for the asymmetry-induced THz emission. Notice that, using Eqs. (\ref{zij}) and (\ref{chi}), by a simple shift in the origin of the $z$ axis, we can shift both $\chi_{11,{q},q_{z}}$ and $\chi_{22,{q},q_{z}}$ by the same amount, whilst keeping $\chi_{12,{q},q_{z}}$ constant. We can thus simplify \Eq{Vthz0} into
\begin{eqnarray}
V\subthz= \sum_{\mathbf{k},\mathbf{q},q_z}&&\Delta\chi_{{q},q_{z}}c^{\dagger}_{2,\mathbf{k+q}}\ca{2}{k}(\aac{-q}{q_z}+\aaa{q}{q_z}),
\label{Vthz}
\end{eqnarray}
where $\Delta\chi_{{q},q_{z}}=\chi_{22,{q},q_{z}}-\chi_{11,{q},q_{z}}$.
Thanks to the fact that $V\subthz$ annihilates the ground state
\begin{eqnarray}
\label{ann}
V\subthz\ket{G}\ket{0_{ph}}&=&0,
\end{eqnarray}
where $\ket{0_{ph}}$ is the vacuum state for the electromagnetic field,
matrix elements of $V\subthz$ between the different eigenstates of \Eq{psi} can thus be easily calculated from the commutators of $V\subthz$. Straightforward algebra gives
\begin{eqnarray}
\label{comm}
\lbrack V\subthz, M_{\mathbf{k}}^{\pm}\rbrack&=&\frac{1}{\sqrt{2}}\sum_{\mathbf{q},q_z} \Delta\chi_{{q},q_{z}}F_{\mathbf{k+q}}E_{\mathbf{k}}(\aac{-q}{q_z}+\aaa{q}{q_z}),\nonumber \\
\lbrack V\subthz, F_{\mathbf{k}}\rbrack&=&\sum_{\mathbf{q},q_z}\Delta\chi_{q,q_{z}}F_{\mathbf{k+q}}(\aac{-q}{q_z}+\aaa{q}{q_z}),\nonumber\\
\lbrack V\subthz, E_{\mathbf{k}}\rbrack&=&0.
\end{eqnarray}
%
\begin{figure}[t!]
\begin{center}
\includegraphics[width=8.6cm]{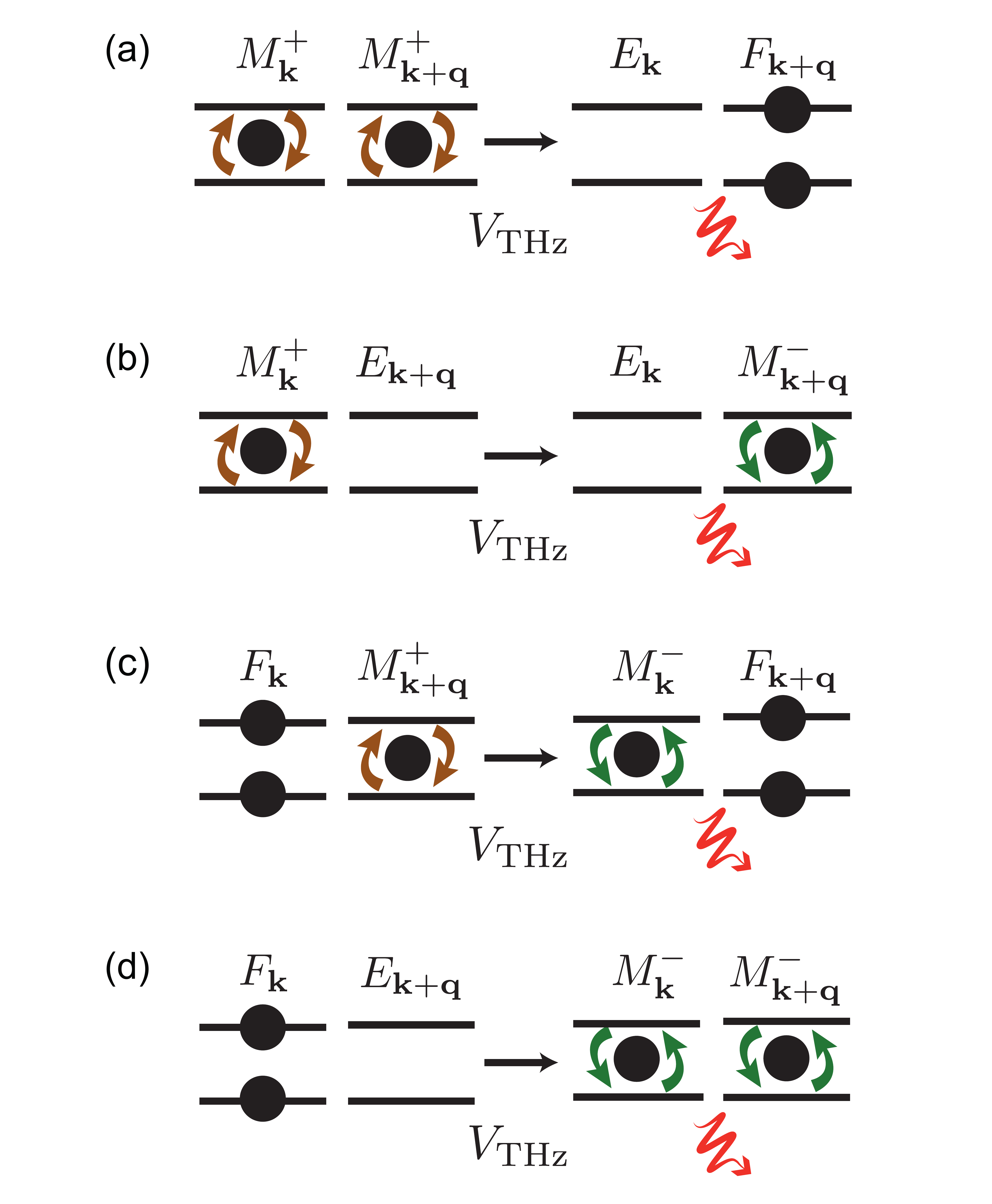}
\caption{
\label{scattering}
(a)-(d): Schematical representation of the four resonant processes involved in the emission of THz photons.}
\end{center}
\end{figure}
\subsection{Emission rates}
\label{Emrates}
We have now all the tools we need to calculate the THz emission due to the asymmetric dipole. 
To this aim we employ Fermi's Golden Rule,
 \begin{eqnarray} 
\Gamma&=&\frac{2\pi}{\hbar^{2}}\sum_{f}|\bra{\psi_{f}}V\subthz\ket{\psi_{i}}|^{2} \delta(\omega_{i}-\omega_{f}),
\label{fgr}
\end{eqnarray}
where the initial and final states, $\ket{\psi_{i}}$ and $\ket{\psi_{f}}$, are eigenstates of $H_{0}'$ with energies $\hbar\omega_{i}$ and $\hbar\omega_{f}$, respectively.
We start by calculating the emission induced by $V\subthz$ in the simplest case in which all the electrons are  cycling between the two subbands under the effect of the pump, and no electrons are blocked in double excitation states ($N_F=0$). This assumption is supported by previous experiments, with limited asymmetry samples \cite{Frogley06,Dynes05, Dynes05B}, where the fraction of electrons participating  in the coherent Rabi oscillation has been estimated to be up to 90$\%$. Although this assumption neglects the electrons that end up in blocked states due to the THz emission, including the latter does not  alter the results significantly. Proving this, however, requires a rather cumbersome algebraic calculation which has been relegated to the Appendix for the sake of simplicity.
 
Using the notation of \Eq{psi}, we are thus considering emission starting from states of the form\begin{eqnarray}
\ket{\psi_i}&=&\prod_{\mathbf{k}\in S_+}M^{+}\sub{k}\prod_{\mathbf{k}\in S_-}M^{-}\sub{k}\ket{G}\ket{0_{ph}},
\label{psii}
\end{eqnarray}
whose energy we will call $\hbar\omega_i$.
The effect of $V\subthz$ can be calculated by commuting it all the way to the right and using Eqs. (\ref{prod}), (\ref{ann}), and (\ref{comm}),
\begin{eqnarray}
\label{Veff}
V\subthz\ket{\psi_i}&=&\frac{1}{\sqrt{2}}\sum\limits_{\substack{\mathbf{k,q},q_z \\ \mathbf{k}\in S_{+}+S_{-}}}
\Delta\chi_{{q},q_{z}}F_{\mathbf{k+q}}E_{\mathbf{k}} \\&\nonumber&\prod\limits\sub{k'\neq k}M^{{j\sub{k'}}}\sub{k'} a^{\dagger}_{\mathbf{-q},q_z} \ket{G}\ket{0_{ph}}\nonumber \\&=&
\sum\limits_{\substack{\mathbf{k,q},q_z \\ \mathbf{k},\mathbf{k+q}\in S_{+}+S_{-}}}
\frac{j\sub{k+q}}{2}\Delta\chi_{{q},q_{z}}F_{\mathbf{k+q}}E_{\mathbf{k}} \nonumber\\&\nonumber&\prod\limits\sub{k'\neq k,k+q}M^{j\sub{k'}}\sub{k'} a^{\dagger}_{\mathbf{-q},q_z}\ket{G}\ket{0_{ph}},\end{eqnarray}
where $j\sub{k}=\pm$, and the sums and products over $\mathbf{k}$, here and in the remainder of the paper, are intended to be for $k<k_{\text{F}}$  unless otherwise specified. In \Eq{Veff} we have neglected border terms, when an electron inside the Fermi sphere is scattered to the outside of it. The latter involve only electrons at a distance $q$ from the border of the Fermi sphere and are thus negligible given that $q/k_F\ll 1$.
The r.h.s. of \Eq{Veff} is a sum of terms that we can recognize, from \Eq{psi}, to be the eigenstates of $H'_0$. These states, 
\begin{eqnarray}
\label{psif}
\ket{\psi_f(\mathbf{k,q},q_z)}&=& F_{\mathbf{k+q}}E_{\mathbf{k}} \prod\limits\sub{k'\neq k,k+q}M^{j\sub{k'}}\sub{k'} a^{\dagger}_{\mathbf{q},q_z} \ket{G}\ket{0_{ph}},\quad\quad
\end{eqnarray}
with energy
\begin{eqnarray}
\label{Ene}
\hbar\omega_f(\mathbf{k,q},q_z)&=&\hbar\omega_{i}+\hbar\omega_{q,q_z}- \frac{\hbar\Omega}{2}\lbrack (j\sub{k}1)+(j\sub{k+q}1)\rbrack,\quad
\end{eqnarray}
will thus be the available final states for the scattering process leading to THz emission.
In particular, in order to satisfy energy conservation, from \Eq{Ene}, the only final states giving rise to a photonic emission will be those with
\begin{eqnarray}
\label{cond}
j\sub{k}=j\sub{k+q}=+.
\end{eqnarray}
At this point we can apply Fermi's Golden Rule to calculate the THz emission rate as
\begin{eqnarray}
\Gamma\subthz=\frac{2\pi}{\hbar^{2}}\sum_{\mathbf{k},\mathbf{q},q_{z}} \frac{|\Delta\chi_{q,q_{z}}|^{2}}{4}\delta({\Omega-\omega_{q,q_{z}}}),\quad\quad
\label{last1}
\end{eqnarray}
from which one can check that the only matrix  elements giving a non-zero contribution are those that respect \Eq{cond}.
As we are assuming that all the electrons are cycling between the two subbands under the effect of the pump laser, on average one quarter of the 
states will respect such a condition.
Given this, and the fact that the terms in the sum of 
Eq. (\ref{last1}) do not depend on $\mathbf{k}$, we can rewrite \Eq{last1} as
\begin{eqnarray}
\Gamma\subthz=\frac{N\pi}{8\hbar^{2}}\sum_{\mathbf{q},q_{z}}|\Delta\chi_{q,q_{z}}|^{2}\delta({\Omega-\omega_{q,q_{z}}}).\quad\quad
\label{last2}
\end{eqnarray}
Note that this result depends only on the average number of electrons in the $M_{\mathbf{k}}^{\pm}$ states, and not on their relative phases. The emission will thus be unaffected by electron dephasing, that would damp the coherence of the collective Rabi oscillations after only a few oscillations \cite{Dynes05}. 
In the following of this paper, we use \Eq{last2} to estimate the quantum efficiency of the THz emission process, both in the case of a free-space emitter and when using a THz cavity to enhance the emission rate.

\section{Quantum efficiency}
\label{Num}

\subsection{Numerical results for an asymmetric QW}
\label{Numsub}

In order to estimate the achievable THz emission rates, and assess the device's technological potentiality, we start here by studying a simple but realistic asymmetric QW structure. The exact degree of asymmetry will have to be chosen carefully; on the one hand, a small degree of asymmetry gives a large $z_{12}$, but a small $\Delta z$, and on the other, excessive asymmetry leads to a large $\Delta z$, but a vanishing $z_{12}$. An efficient THz emitter design must lie between these extremes, with an acceptably large intersubband dipole $ez_{12}$, in order to couple strongly to the pump beam, and a sizeable asymmetric dipole $e\Delta z$, to efficiently emit THz radiation.

We consider an infinite double QW structure, as shown in Fig. \ref{isbt+qw}(a), comprising a central barrier of height $V_{0}$ and width $L_{b}$,  separating two potential wells of widths $L_{a}$ and $L_{c}$, respectively. The overall QW, of total length $L_{QW}$, is asymmetric for $L_{a}\neq  L_{c}$. 
\begin{figure*}[t!]
\begin{center}
\includegraphics[width=12.6cm]{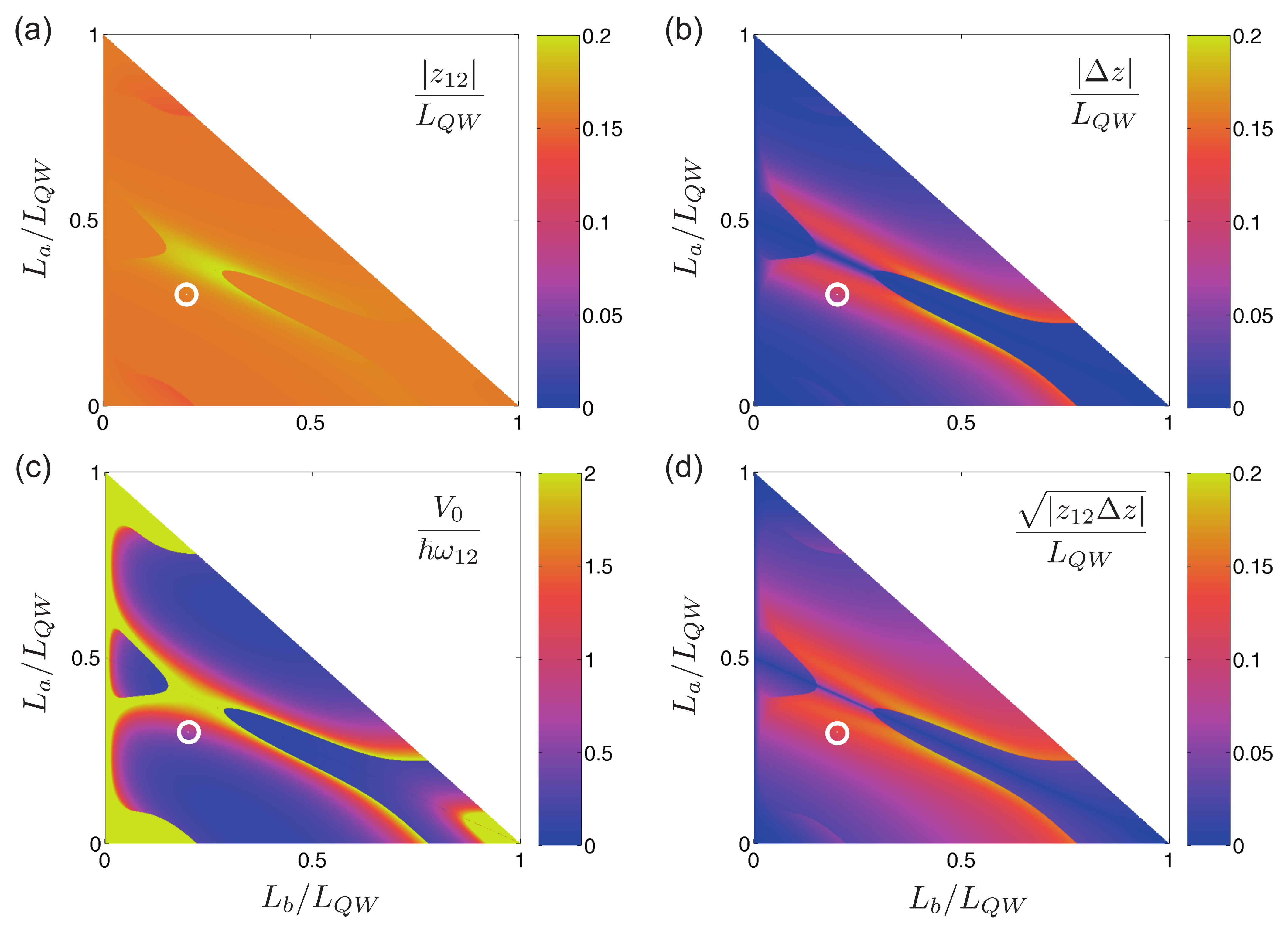}
\caption{Intersubband (a) and asymmetric (b) dipoles, obtained using the optimization procedure described in the text. In panels (c) and (d) we plot the optimal value of $V_0$ and of $|z_{12}\Delta z|$, respectively. A circle highlights the point in the parameter space that is used for the numerical estimates given in the main text.
\label{numerical}}
\end{center}
\end{figure*}
We explore the parameter space by varying $L_{a}$, $L_{b}$, and $V_{0}$, in order to maximize the value of $|z_{12}\Delta z|$. This figure of merit allows us to identify structures with sizeable values for both dipoles and, as we will see, the quantum efficiency of the THz emission explicitly depends upon this parameter. The optimization is performed keeping $L_{QW}$ and $\hbar\omega_{12}$ fixed. This procedure thus mimics the search for an optimal structure given a fixed QW's length and pump laser frequency. In particular, we set $\hbar\omega_{12}=125$ meV and $L_{QW}=11.6\ $ nm, which is the length corresponding to the desired $\hbar\omega_{12}$ when $V_0=0$. The height of the central barrier $V_0$ is allowed to vary up to $250\ $meV, mindful of the barrier heights obtainable in Al$_x$Ga$_{1-x}$As heterostructures.  
The panels (a) and (b) of Fig. \ref{numerical},  show dipole values as a function of $L_{b}$ and $L_{a}$, in units of $L_{QW}$. The corresponding height of the barrier, $V_{0}$, and the value of the maximized figure of merit (normalized to facilitate comparison), are shown in panels (c) and (d). As expected, the asymmetric dipole $\Delta z$ vanishes on the line $L_{a}/L_{QW}=\tfrac{1}{2}(1-L_{b}/L_{QW})$, corresponding to a symmetric QW case. 

For the sake of definiteness, and in order to make numerical estimates, henceforth we set the parameters $L_{a}=0.3L_{QW}$, $L_{b}=0.2L_{QW}$, denoted by the circled point in the parameter space in Fig. \ref{numerical}, leading to  $z_{12}=0.18L_{QW}$ and $\Delta z=0.11L_{QW}$. While larger values for the dipoles are  obtainable in principle, we have chosen these values because they generate results that are stable over a fairly large section of the parameter space, making them robust against device fabrication tolerances. 
It is important to notice that these dipole values are almost one order of magnitude larger than the values obtainable in quantum dots, for similar emission frequencies \cite{Kibis09}. 

\subsection{Free-space emission}

The free-space THz emission rate $\Gamma\subthz^{0}$ of a single QW, per unit surface $S$, can be estimated directly from \Eq{last2} with the expressions given by Eqs. (\ref{chi}) and (\ref{zpf}), by transforming the sum over discrete photonic states into an integral,
 \begin{eqnarray}
\frac{\Gamma^{0}\subthz}{S}
&=&\frac{N_{2DEG}e^2|\Delta z|^2 \sqrt{\epsilon_r}}{12  \epsilon_0 \pi\hbar c^3}\Omega^3,
 \label{gthz}
 \end{eqnarray}
where we considered a QW doped with a uniform surface density $N_{2DEG}=N/S$. 
This formula has the same parameter dependency as the emission formula developed in Ref. \cite{Kibis09} for emission from quantum dots. From \Eq{gthz}, and from the fact that both the asymmetric dipole $\Delta z$ and the dipole density $N_{2DEG}$ of the structure we are considering are one order of magnitude larger than in the quantum dots case, we can expect an emission rate three orders of magnitude larger than in previous quantum dot-based proposals. A posteriori this confirms that QWs are an ideal testbed to observe intra-doublet emission.

\begin{figure}[t!]
\begin{center}
\includegraphics[width=8.6cm]{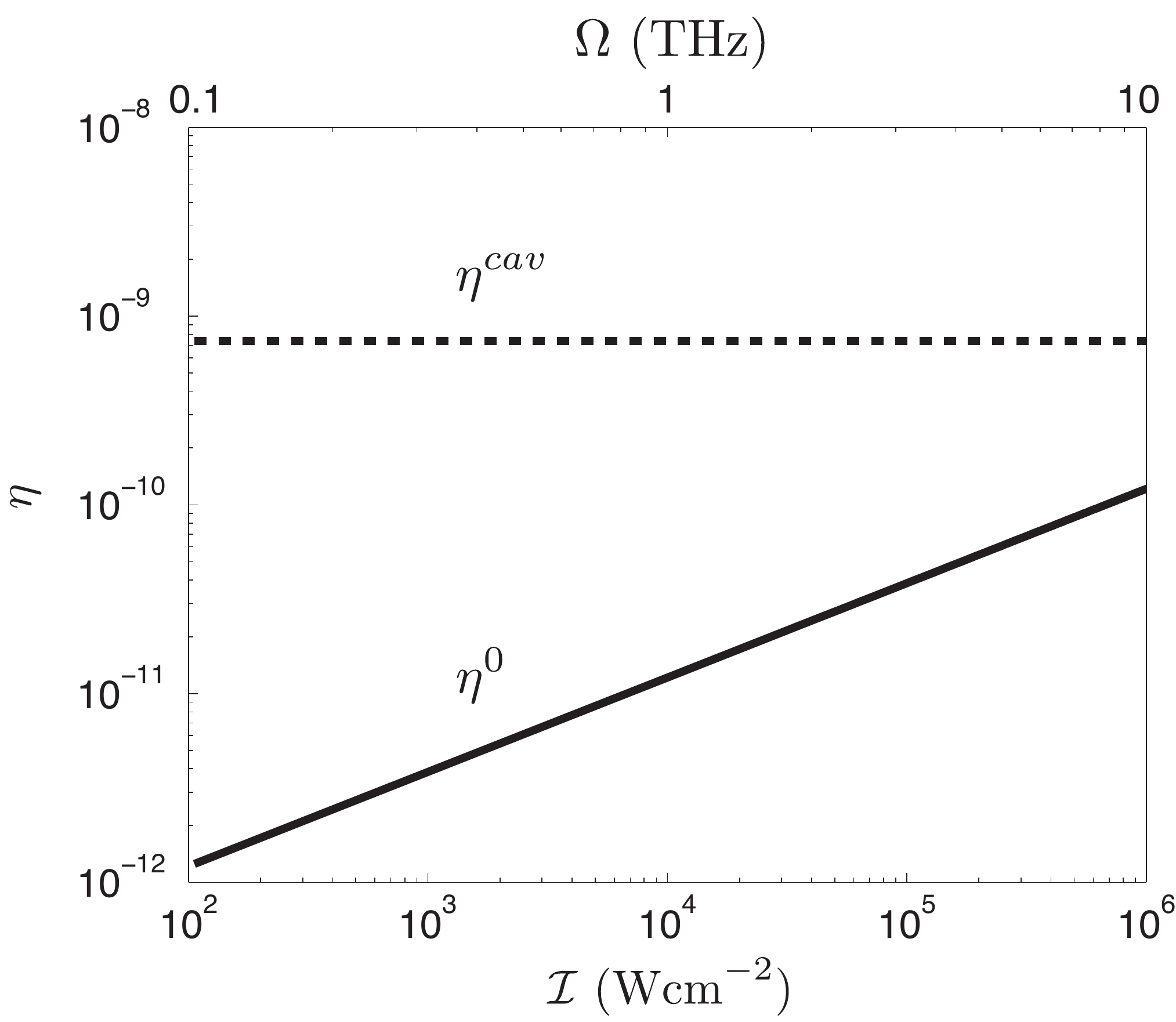}
\caption{\label{etafig}
Quantum efficiency for free space $\eta^0$ (solid line) and cavity emission $\eta^{cav}$ (dashed line), as a function of the pump power density $\mathcal{I}$ (lower axis) and of the Rabi frequency $\Omega$ (upper axis). }
\end{center}
\end{figure}

The free-space quantum efficiency is then given by
\begin{eqnarray}
\eta^{0}&=&\frac{\hbar\omega_{12}N_{QW}\Gamma\subthz^0}{\mathcal{I}}\\\nonumber &=&\frac{N_{QW}N_{2DEG}e^4|\Delta z|^2 z_{12}^2 \sqrt{\epsilon_{r}} \omega_{12}\Omega}{6 \epsilon_{0}^2 \pi \hbar^2 c^4 },
\label{eta0}
\end{eqnarray}
where the device is made of $N_{QW}$ identical QWs  and $\mathcal{I}=\epsilon_{0}c\mathcal{E}^{2}/2$ is the pump power density.
The free-space efficiency for $N_{QW}=50$ and $N_{2DEG}= 10^{12}$ cm$^{-2}$, and using the QW whose parameters are marked by a white circle in Fig. \ref{numerical}, is plotted in Fig. \ref{etafig} (solid line) as a function of the the pump power density, $\mathcal{I}$ (lower horizontal axis), and of the emitted frequency $\Omega$ (upper horizontal axis).
In particular, considering  a pump of strength $\mathcal{I}=8\cdot10^{6}$ Wcm$^{-2}$, from \Eq{ez12} one obtains $\Omega/2\pi=1$ THz, which is of the same order of those experimentally observed in Ref. \cite{Dynes05}. The free-space quantum efficiency would be then $\eta^{0}\simeq 10^{-10}$.

\subsection{Cavity emission rate}
\label{Cav}
In order to increase the THz emission rate, it is possible to embed the multiple-QW structure into a THz cavity \cite{Kavokin}. As an analysis of different kinds of THz resonators is out of the scope of the present work, we will limit ourselves to the conceptually simple case of a planar THz cavity. A modification of \Eq{gthz}, considering a two-dimensional continuum of photonic modes then gives \cite{Kakazu94}
\begin{eqnarray}
\frac{\Gamma\subthz^{cav}}{S}&=&\frac{N_{2DEG}e^2|\Delta z|^2}{32 \hbar\epsilon_0c^{2}L_{cav}}\Omega^{2},
\label{gthzcav}
\end{eqnarray} 
where $L_{cav}$ is the cavity length. 
The quantum efficiency accordingly becomes
\begin{eqnarray}
\eta^{cav}&=&\frac{\hbar\omega_{12}N_{QW}\Gamma\subthz^{cav}}{\mathcal{I}}\\\nonumber &=&\frac{N_{QW}N_{2DEG}e^4|\Delta z|^2 z_{12}^2\omega_{12}}{16 \epsilon_{0}^2 \hbar^2 c^3 L_{cav} },
\label{eta0}
\end{eqnarray}
independent of the pump power in the parameter regime we are considering.
Note that both $\eta^0$ and $\eta^{cav}$ are  proportional to $|z_{12} \Delta z|^2$ (the former only at given emission frequency). This justifies a posteriori our choice of using $|z_{12} \Delta z|$ as our optimisation parameter in Sec. \ref{Numsub}.
The efficiency gain using a two dimensional cavity is thus given by the Purcell factor \cite{Purcell46}, $F_P=\Gamma^{cav}\subthz/\Gamma^0\subthz$, i.e.
\begin{eqnarray}
F_P=\frac{3\pi c}{8\sqrt{\epsilon_r}L_{cav}\Omega}.
\label{fp}
\end{eqnarray}
To increase the emission efficiency at a fixed pump power it is thus convenient to reduce the cavity length $L_{cav}$. Present-day THz cavities allow for strong sub-wavelength confinement \cite{Todorov07}, using plasmonic or localized phonon-plasmon excitations \cite{Palma12,Caldwell13}. These cavities have demonstrated a linear confinement $\lambda_{res}/L_{cav}\simeq 200$, 
where $\lambda_{res}$ is the free-space wavelength of the THz radiation, with quality factors in excess of $100$. 
The cavity efficiency, $\eta^{cav}$, is independent on the pump strength, as shown in Fig. \ref{etafig} (dashed line); for the parameters chosen in the previous section, and a cavity length $L_{cav}=1\ \mu$m, we obtain $\eta^{cav}\simeq 10^{-9}$,
which is competitive with  fluorescence efficiency in monolithic THz emitters, with tunable frequencies \cite{Kibis09}.
 
\section{Conclusions}
\label{Conclusions}
We have shown how asymmetric artificial atoms can be exploited to obtain a resonant fluorescence THz peak from a  transition that would normally be dipole-forbidden in centro-symmetric systems. We have developed a many-body theory that allows us to give a reliable estimate of the photon emission achievable in a realistic device, showing that the emission rate can be orders of magnitude larger than in previous quantum dot-based proposals. Numerical results of the attainable efficiency indicate that such an emission channel should be observable in present day experiments, and it could potentially be exploited to realize monolithic THz devices.   

\section{Acknowledgments}
We thank  P. G. Lagoudakis, A. V. Kavokin, and S. Portolan for fruitful discussions.
SDL is Royal Society Research Fellow. SDL and NS acknowledge support from the Engineering and Physical Sciences Research Council (EPSRC), research grant EP/L020335/1.

\appendix

\section{Emission from general eigenstates}
\label{AppendixA}
In Sec. \ref{Emrates}, we calculated the THz emission, limiting ourselves to the case in which
no electrons are locked into double-occupancy states ($N_F=0$). However, under the effect of the pump beam, electrons will be actually scattered into such states, as shown in Fig. \ref{scattering}(c). Other scattering mechanisms, both radiative and non-radiative, will subsequently scatter away those blocked electrons. In this Appendix we develop a more detailed theory taking into account these processes, considering the emission for a general state in the form of \Eq{psi}. 

Applying $V\subthz$ to such a state we obtain, using Eqs. (\ref{prod}), (\ref{ann}) and (\ref{comm})
\begin{widetext}
\begin{eqnarray}
\label{Veffg}
V\subthz\ket{\psi_i}&=&\frac{1}{\sqrt{2}}\sum\limits_{\substack{\mathbf{k},\mathbf{q},q_z\\ \mathbf{k}\in S_+}}
\Delta\chi_{{q},q_{z}}F_{\mathbf{k+q}}E_{\mathbf{k}}\nonumber\prod_{\mathbf{k'}\in S_+ + S_- -\{\mathbf{k}\}}M^{j\sub{k'}}\sub{k'} \prod_{\mathbf{k'}\in S_F}F\sub{k'}
\prod_{\mathbf{k'}\in S_E}E\sub{k'} a^{\dagger}_{\mathbf{-q},q_z} \ket{G}\ket{0_{ph}}  \\&+&
\sum\limits_{\substack{\mathbf{k},\mathbf{q},q_z\\ \mathbf{k}\in S_F}}\Delta\chi_{{q},q_{z}}
F_{\mathbf{k+q}}  \prod_{\mathbf{k'}\in S_+}M^+\sub{k'} \prod_{\mathbf{k'}\in S_-}M^-\sub{k'}
   \prod_{\mathbf{k'}\in S_F-\{\mathbf{k}\}}F\sub{k'}
\prod_{\mathbf{k'}\in S_E}E\sub{k'}  
a^{\dagger}_{\mathbf{-q},q_z} \ket{G}\ket{0_{ph}}.
\end{eqnarray}
\end{widetext}
The two lines of  \Eq{Veffg} give rise to $4$ different terms each, depending on which set $\mathbf{k+q}$ belongs to, each of these terms describing a different scattering channel.
At the same time, only 4 of these 8 terms, those for which the difference between the numbers of $M^+_{\mathbf{k}}$ and $M^-_{\mathbf{k}}$ operators is strictly smaller than $N_+-N_-$, will give rise to resonant emission processes.
Moreover, we make the assumption, to be confirmed a posteriori, that most of the electrons are coupled to the laser pump, and only few are locked in double-occupancy $F_{\mathbf{k}}$ states. We can thus limit ourselves to terms of the lowest order in $\frac{N_F}{N}$.

In the first line of \Eq{Veffg}, if both $\mathbf{k}$ and $\mathbf{k+q}$ are in $S_+$, we obtain a result analogous to that of the previous section, describing a scattering from two single-occupancy states to a full and and an empty state, the only difference being the normalization of the sum over electronic and photonic wavevectors in \Eq{last1}. The total contribution to the emission of the scattering process from states such that $\mathbf{k,k+q}\in S_+$, sketched in Fig. \ref{scattering}(a), can be estimated by using \Eq{Ntot}, assuming that $N_{+}=N_{-}$, so that after straightforward manipulation, $\frac{N_+^2}{N}\simeq (\frac{N}{4}-N_F)$, where the term of order $N_{F}^{2}/N$ has been neglected. We thus obtain
\begin{eqnarray}
\Gamma_{THz}^{++\rightarrow EF}&=&(1-4\frac{N_F}{N})\Gamma_{THz}.
\end{eqnarray}
If instead  $\mathbf{k+q}$ belongs to $S_E$, we obtain 
\begin{widetext}
\begin{eqnarray}
\label{Veffg2}
\frac{1}{\sqrt{2}}
\sum\limits_{\substack{\mathbf{k},\mathbf{q},q_z\\ \mathbf{k}\in S_+,\mathbf{k+q}\in S_E}}
\Delta\chi_{{q},q_{z}}F_{\mathbf{k+q}}E_{\mathbf{k}}\nonumber\prod_{\mathbf{k'}\in S_++S_--\{\mathbf{k}\}}M^{j\sub{k'}}\sub{k'} \prod_{\mathbf{k'}\in S_F}F\sub{k'}
\prod_{\mathbf{k'}\in S_E}E\sub{k'} a^{\dagger}_{\mathbf{-q},q_z} \ket{G}\ket{0_{ph}}  =\\
\frac{1}{{2}}\sum\limits_{\substack{\mathbf{k},\mathbf{q},q_z\\ \mathbf{k}\in S_+,\mathbf{k+q}\in S_E}}
\Delta\chi_{{q},q_{z}}(M^+_{\mathbf{k+q}}+M^-_{\mathbf{k+q}})E_{\mathbf{k}}\prod_{\mathbf{k'}\in S_++S_--\{\mathbf{k}\}}M^{j\sub{k'}}\sub{k'} \prod_{\mathbf{k'}\in S_F}F\sub{k'}
\prod_{\mathbf{k'}\in S_E-\{\mathbf{k+q}\}}E\sub{k'} a^{\dagger}_{\mathbf{-q},q_z} \ket{G}\ket{0_{ph}},
\end{eqnarray}
describing a process, sketched in Fig. \ref{scattering}(b), in which an electron with energy $\hbar\Omega$ scatters into an empty $\mathbf{k}$-subspace, giving rise to a state of energy $-\hbar\Omega$, an empty one, and a photon. The emission rate of this process can be calculated analogously to what has been done in \Eq{last2}, but with the normalization $\frac{N_+N_F}{N}$
\begin{eqnarray}
\Gamma_{THz}^{+E\rightarrow E-}&=&\frac{2N_F}{N}\Gamma_{THz}.
\end{eqnarray}
Finally, the second line of \Eq{Veffg} gives a non-negligible contribution only for $\mathbf{k+q}\in S_+$, describing a process, sketched in Fig. \ref{scattering}(c),  in which one of the two electrons of a full state, and an electron in a state with energy $\hbar\Omega$, scatter into an full state and a state with energy $-\hbar\Omega$, and we obtain
\begin{eqnarray}
\label{Veffg3}
\sum\limits_{\substack{\mathbf{k},\mathbf{q},q_z\\ \mathbf{k}\in S_F,\mathbf{k+q}\in S_+}}
F_{\mathbf{k+q}}  \prod_{\mathbf{k'}\in S_+}M^+\sub{k'} \prod_{\mathbf{k'}\in S_-}M^-\sub{k'}
   \prod_{\mathbf{k'}\in S_F-\{\mathbf{k}\}}F\sub{k'}
\prod_{\mathbf{k'}\in S_E}E\sub{k'}  
a^{\dagger}_{\mathbf{-q},q_z} \ket{G}\ket{0_{ph}}=\quad\quad\\\nonumber
\frac{1}{\sqrt{2}}\sum\limits_{\substack{\mathbf{k},\mathbf{q},q_z\\ \mathbf{k}\in S_F,\mathbf{k+q}\in S_+}}\Delta\chi_{{q},q_{z}}
F\sub{k+q} \prod_{\mathbf{k'}\in S_+-\{\mathbf{k+q}\}}M^+\sub{k'} \prod_{\mathbf{k'}\in S_-}M^-\sub{k'}
   \prod_{\mathbf{k'}\in S_F-\{\mathbf{k}\}}F\sub{k'}
\prod_{\mathbf{k'}\in S_E}E\sub{k'}  
a^{\dagger}_{\mathbf{-q},q_z} \ket{G}\ket{0_{ph}}.
\end{eqnarray}
\end{widetext}
Also in this case the relevant normalization for the sum over the wavevectors is $\frac{N_+N_F}{N}$, giving an emission rate
\begin{eqnarray}
\Gamma_{THz}^{F+\rightarrow -F}&=&\frac{2N_F}{N}\Gamma_{THz}.
\end{eqnarray}
The remaining emission process, $FE\rightarrow --$, sketched in Fig. \ref{scattering}(e), can be ignored, as the scattering occurs from two initial uncoupled states, $F_{\mathbf{k}}$ and $E_{\mathbf{k+q}}$, and it is thus of second order in $N_{F}$. 

We thus obtain the important result that, to the first order in $N_F$, the emission rate does not depend on $N_F$, as
\begin{eqnarray}
\Gamma_{THz}^{++\rightarrow EF}+\Gamma_{THz}^{+E\rightarrow E-}+\Gamma_{THz}^{F+\rightarrow -F}=\Gamma_{THz}.
\end{eqnarray}
In order to ascertain if the first order approximation is adequate, we need to estimate the number of electrons that are locked in double occupancy states. In fact, the semiconductor also allows many non-radiative energy relaxation channels that cool the electron gas, so we can write, always to the first order in $N_F$, a rate equation for the total number of electrons that are coupled to the pump laser $N_e=N_++N_-$,
\begin{eqnarray}
\label{pop}
\frac{dN_{e}}{dt}=-2\Gamma^{++\rightarrow EF}\subthz +2\Gamma_{nr}N_F,
\end{eqnarray}  
where $\Gamma_{nr}$ is the rate of non-radiative relaxation. At equilibrium we have
\begin{eqnarray}
\label{popeq}
\frac{N_F}{N}&=&\frac{1}{4+\frac{N\Gamma_{nr}}{\Gamma\subthz}}.
\end{eqnarray}
Given the extremely fast non-radiative phonon-assisted intersubband recombination, $N_F$ in \Eq{popeq} can be safely taken to be vanishing, thus a posteriori confirming our initial approximation.

\end{document}